\documentclass[sigconf]{acmart}
 
\AtBeginDocument{%
  \providecommand\BibTeX{{%
    \normalfont B\kern-0.5em{\scshape i\kern-0.25em b}\kern-0.8em\TeX}}}

\copyrightyear{2027}
\acmYear{2027}
\setcopyright{cc}
\setcctype{by}
\acmConference[GROUP Companion '27]{The 2027 ACM International Conference on Supporting Group Work}{January 10--13, 2027}{St. Simons Island, GA, USA}
\acmBooktitle{The 2027 ACM International Conference on Supporting Group Work (GROUP Companion '27), January 10--13, 2027, St. Simons Island, GA, USA}
\acmDOI{10.1145/3838685.3849858}
\acmISBN{979-8-4007-2916-4/2027/01}

\usepackage{booktabs} 
\usepackage{url}

\usepackage{caption}
\usepackage[all]{nowidow}
\usepackage{array}
\usepackage{arydshln}
\usepackage{tabularx}
\usepackage{multirow}
\usepackage{arydshln}
\newcolumntype{L}[1]{>{\raggedright\let\newline\\\arraybackslash\hspace{0pt}}m{#1}}
\newcolumntype{C}[1]{>{\centering\let\newline\\\arraybackslash\hspace{0pt}}m{#1}}
\newcolumntype{R}[1]{>{\raggedleft\let\newline\\\arraybackslash\hspace{0pt}}m{#1}}

\usepackage{pgfplots}
\pgfplotsset{compat=1.18}

\def\authnotes{1}
\newcounter{notectr}[section]
\newcommand{\thenote}{\thesubsection.\arabic{notectr}\refstepcounter{notectr}}

\newcommand{\note}[2]{$\ll$#1~\thenote: #2$\gg$}
\newcommand{\cnote}[1]{\ifnum\authnotes=1 \textcolor{blue}{\note{Comment:}{#1}}\fi}

\begin{document}


\title[Navigating Reproductive Well-being Conversations with Young Adults]{ ‘OpenBloom': A Stigma-Sensitive LLM Design Probe for Navigating Reproductive Well-being Conversations with Young Adults}
 
\author{Yang Hong}
\authornote{These authors contributed equally to this work.}
\affiliation{%
  \institution{ University of Illinois Urbana-Champaign}
  \city{Champaign}
  \state{Illinois}
  \country{United States}}
\email{yangh9@illinois.edu}

\author{Ashley Hua}
\authornotemark[1]
\affiliation{%
  \institution{University of Illinois Urbana-Champaign}
  \city{Champaign}
  \state{Illinois}
  \country{United States}}
\email{ahhua2@illinois.edu}

\author{Adya Daruka}
\authornotemark[1]
\affiliation{%
  \institution{University of Illinois Urbana-Champaign}
  \city{Champaign}
  \state{Illinois}
  \country{United States}}
\email{adaruka2@illinois.edu}

\author{Sharifa Sultana}
\affiliation{%
  \institution{ University of Illinois Urbana-Champaign}
  \city{Champaign}
  \state{Illinois}
  \country{United States}}
\email{sharifas@illinois.edu}

\renewcommand{\shortauthors}{Hong et al.}

\begin{abstract}
The growing use of large language models (LLMs) by young adults seeking sensitive health information has raised important questions in Human-AI Interaction about how these systems can support understanding and navigation of reproductive well-being. In response to Feminist HCI principles, we introduce OpenBloom, a web application and an exploratory design probe that uses LLMs to generate question-based prompts from reproductive health articles. Through a user study with 34 young adults across 136 interactions with OpenBloom, we provide an initial assessment of the system while exploring how participants' reflections engage with culture and value sensitivities. We found that while OpenBloom outputs were generally perceived as "safe" and non-offensive, they tend to paraphrase or rely on factual recall, which may lead to value dilution. We discuss implications under contestability and value-sensitive frameworks for future LLM-mediated reproductive health technologies and towards responsible AI discourse.
\end{abstract}







\settopmatter{printfolios=true}

\maketitle

\section{Introduction}
Reproductive well-being, including menstruation, contraception, fertility, and sexual health, remains stigmatized across many communities~\cite{chowdhury2024ancient, bagalkot2022embodied, chopra2021living}. For young adults, especially young women, reproductive health knowledge is often difficult to access or discuss openly due to cultural taboos that stigmatize these topics as ``forbidden'' conversations~\cite{hartley2023words, alkhalili2024assessment, mohdtohit2024forbidden}. The stigmatization manifests in everyday discourse and affect, such as silence around menstruation in schools, shame surrounding contraceptive use, and discomfort discussing sexual health openly~\cite{leblanc2024breaking, alkhalili2024assessment, mohdtohit2024forbidden}. Beyond personal experiences, stigma of reproductive well-being can also structurally constrain accessibility to healthcare; for example, unmarried women may avoid reproductive health services due to fear of social judgment and discrimination~\cite{mohammadi2016stigma}. Despite sustained public health efforts, challenges persist in how young adults interpret and communicate reproductive health information under stigmatized contexts.

The growing use of large language models (LLMs) in healthcare highlights their potential to support young adults seeking and learning sensitive health information, especially for young women in underserved settings~\cite{Fetrati2025Leveraging, mina2025aichatbots, koulouri2022chatbots, yadav2019feedpal}. Although LLM-based chatbots, positioned as mediators of reproductive health information, are useful in generating explanations, prompts, and questions that adapt to user-provided content~\cite{Deva2025Kya, Fetrati2025Leveraging, yadav2019feedpal}, it remains underexplored how young adults perceive and respond to LLM-generated reproductive health content when its language, assumptions, and values encounter stigma and culture sensitivities.

In this initial study, we examined how young adults interact with and perceive stigma, cultural norms, and the boundaries of what is considered ``safe'' in AI-generated reproductive health prompts. Drawing on the Feminist HCI agenda~\cite{bardzell2010feminist} and previous studies on stigma-sensitive reproductive design~\cite{sultana2025socheton, Deva2025Kya}, we approach the LLM tool not as a finalized solution to reducing health stigma, but as an explorative design probe inviting participation, exposing values and assumptions, and informing future design. To achieve this, we developed \textit{OpenBloom}, a web application built with Flutter and Dart that uses LLMs to transform user-submitted reproductive health articles into prompt questions to initiate relevant conversations. We invited 34 young adults to interact with OpenBloom and provide feedback on their perceived qualities and concerns based on their experience. We found that while participants generally perceived the LLM-generated questions as relevant, correct, and non-offensive, the questions often defaulted to superficial rephrasing or factual recall, lacked critical or reflective depth, constraining further engagement with stigma-sensitive reproductive topics. Participants also emphasized the need for more contextually and culturally nuanced, and even mildly provocative interactions to prompt reflections.

Our study aims to make twofold contributions to reproductive well-being and Human-AI Interaction: (1) We present OpenBloom as a design probe that surfaces the values, tensions, and limitations involved in applying LLMs to reproductive health contexts. (2) We use the findings to inform future work on contestable and value-sensitive approaches to LLM-mediated reproductive health technologies, addressing the risk of value dilution over time and moving toward more reflective, responsible, and socially grounded AI futures. Given GROUP’s interdisciplinary community, we hope to present this poster as an opportunity to gather valuable feedback on how this design probe might be further developed and applied across diverse social and cultural contexts of reproductive well-being.
\section{Background and Design Motivations}

Educational approaches to sensitive health topics, such as reproductive well-being, have increasingly emphasized inquiry-based and question-driven pedagogies to foster critical thinking and personal reflection~\cite{burns2016socratic, polyzois2010problem}. In reproductive health education, such approaches have been shown to improve decision-making and reduce discomfort in discussing stigmatized topics~\cite{millanzi2022effect}. While recent advances in automatic question generation (AQG) and LLMs offer new opportunities to scale question-based learning~\cite{wang2024exploring}, emerging evidence suggests that AI-generated questions tend to emphasize factual recall rather than supporting deeper reflection on stigma, identity, and social norms~\cite{burns2024generativeai, hanai2024generative}. Research in HCI and public health has shown that reproductive health stigma operates across multiple levels, including internalized shame, interpersonal judgment, and structural constraints on access to care~\cite{chowdhury2024ancient, bohren2022strategies}. Digital systems may inadvertently reproduce these dynamics by privileging dominant cultural norms or sanitizing sensitive topics. In AI-mediated contexts, these dynamics are further complicated by the sociotechnical nature of LLMs. Prior studies reveal that LLMs may generate culturally inappropriate or overly generalized responses, reinforce Western-centric assumptions, or produce “sanitized” content that avoids engaging with the social and emotional complexities of reproductive health~\cite{jo2023understanding, Deva2025Kya, omar2025evaluating, hanna2025assessing, bouguettaya2025racial, agrawal2024fairness, omar2025chatbias}.

In response, HCI scholarship has advanced a range of design approaches aimed at countering stigma and supporting more inclusive and culturally responsive forms of learning. Feminist HCI and justice-oriented design frameworks emphasize pluralism, participation, and reflexivity, advocating for systems that center marginalized perspectives and enable users to critically engage with dominant narratives~\cite{bardzell2010feminist, kumar2020taking, almeida2020woman}. Within reproductive health, culturally grounded educational tools and participatory design processes have further demonstrated the importance of contextual sensitivity and community engagement~\cite{tuli2018menstrupedia, nkabane2024nthabi}. Translating these commitments into AI-driven systems remains challenging, as scalability can come at the cost of contextual nuance~\cite{davies2024culturalsensitivity, nadarzynski2024equity}. Although LLMs enable adaptive and personalized reproductive health materials, they may also reproduce dominant norms and limit conversations to surface-level engagement. This tension raises a broader design question of how human values can be incorporated into LLM systems and sustained as those systems are put into use.

Value Sensitive Design (VSD) provides a complementary framework for addressing this question by integrating human values throughout conceptual, empirical, and technical inquiry~\cite{Jia2026Designing, cruzmartinez2021value,jongsma2020valuesensitive}. In healthcare, VSD has been used to connect values such as agency, privacy, trust, empathy, and inclusion to concrete design decisions within socio-technical contexts~\cite{Jia2026Designing,jongsma2020valuesensitive}. Values embedded during design can also shift when technologies enter practice in real-life contexts over time~\cite{Ghoshal2023dilution}. Drawing on Feminist HCI and VSD, we treat stigma sensitivity as a situated and evolving design concern and use OpenBloom as a design probe to explore how intended values are expressed through LLM-generated questions and interpreted by young adults.

\subsubsection*{\textbf{Design Goals}}
Building on prior work in stigma-sensitive design in reproductive well-being~\cite{sultana2025socheton, Deva2025Kya} and value-sensitive design in healthcare~\cite{Jia2026Designing, cruzmartinez2021value, jongsma2020valuesensitive}, OpenBloom was designed with four goals: (1) \textit{supporting reflective and inquiry-based engagement}, by generating questions that encourage critical thinking, reflection, and exploration of sociocultural norms; (2) \textit{reducing defensiveness through empathetic and non-judgmental framing}, by leveraging question-based interactions that acknowledge existing beliefs and promote dialogue; (3) \textit{ensuring cultural sensitivity and inclusive representation}, by avoiding culturally inappropriate assumptions and incorporating diverse identities, experiences, and contexts; and (4) \textit{positioning AI as a supportive mediator}, by integrating human values and oversight and treating the system as a probe that surfaces problems in interactions rather than resolving them. These goals guide the design of OpenBloom as a system that facilitates stigma-aware engagement in reproductive well-being.
\section{Designing OpenBloom}

\begin{figure}[t!]
    \centering
    \includegraphics[width=\linewidth]{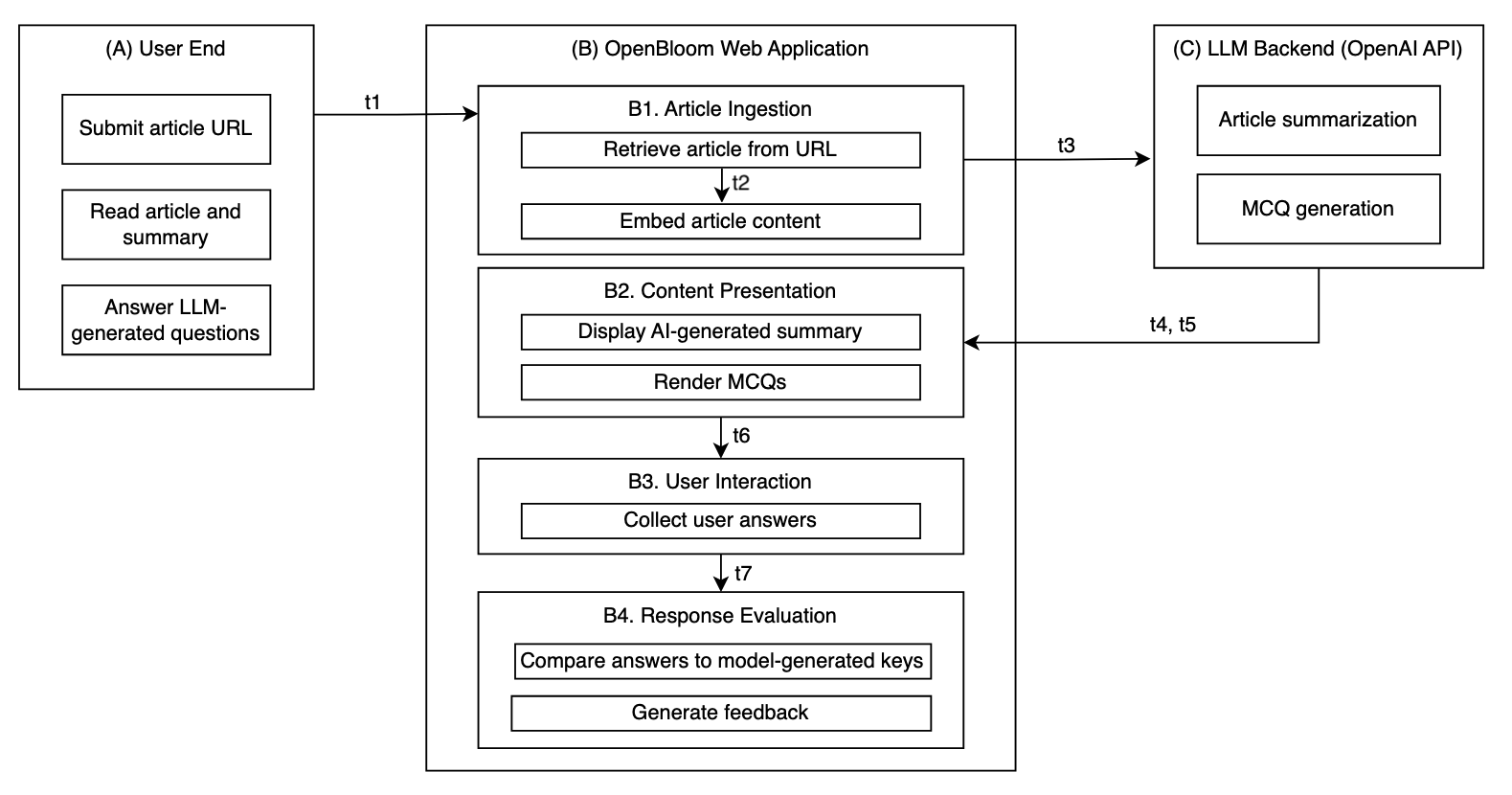}
    \caption{Workflow of OpenBloom. The process begins when the user submits a reproductive health-related article URL to the web application (t1), after which the system retrieves and embeds the article content (t2) and sends the article text to the LLM backend (t3). The LLM then generates a summary (t4) and stigma-related multiple-choice questions (MCQs) (t5), which are presented to the user for interaction (t6). Finally, it compares the user’s responses and model-generated answers and provides feedback (t7).}
    \Description{A workflow diagram of OpenBloom with three components: the user end, the OpenBloom web application, and the LLM backend. The user submits an article URL. OpenBloom retrieves and embeds the article, sends the content to the LLM for summarization and multiple-choice question generation, presents the generated content to the user, collects responses, compares them with model-generated answers, and provides feedback.}
    \label{fig:workflow}
\end{figure}

\begin{figure}[t!]
    \centering
    \includegraphics[width=\linewidth]{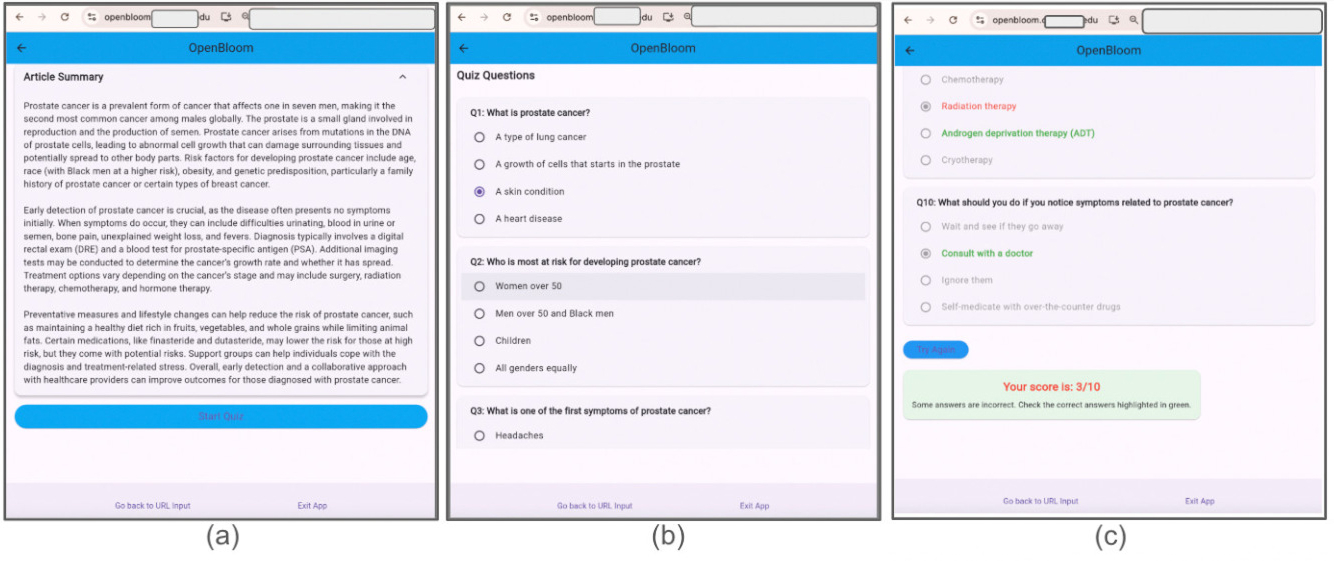}
    \caption{(a) Article summary generated by OpenBloom, (b) excerpt of questions generated by OpenBloom, and (c) correct answer displayed after the participant submits a response.}
    \Description{Three screenshots of the OpenBloom interface. Panel (a) shows an AI-generated summary of a reproductive health article. Panel (b) shows multiple-choice questions generated from the article. Panel (c) shows the user's submitted answers, highlights the correct responses, and displays the resulting score.}
    \label{fig:landing}
\end{figure}

Inspired by previous work that integrates reproductive care with LLM applications \cite{sultana2025socheton, antoniak2024nlp, jo2023understanding}, we developed a website application ``OpenBloom'' using Flutter and Dart as the data collection platform. We chose Flutter for its cross-platform compatibility and Dart for its efficient handling of asynchronous operations required for OpenAI API calls, enabling rapid prototyping on Android devices. The application workflow begins with user input: users choose and submit a URL to a reproductive health-related article (t1). The app embeds the full article within the interface (t2) and then uses OpenAI's GPT-4o model to generate a concise, accessible summary (t3--t4). Using the article as context, the LLM generates a series of questions to promote engagement with the topic (t5--t7). The labels t1–t7 denote the corresponding system features illustrated in Figure~\ref{fig:workflow}. These questions emphasize key points and encourage reflections on issues related to reproductive literacy, social perceptions, and relevant stigmas. Users answer the questions in the application (Figure~\ref{fig:landing}) and receive immediate feedback from OpenBloom. It allows us to collect real-time data on how users engage with AI-generated questions across various reproductive well-being topics, and provides insights into culture and value sensitivities of LLM outputs in highly stigmatized domains.

\section{Collecting Users' Feedback}

We evaluated OpenBloom with 34 young adults (27 women and 7 men), aged 18–29 years (M = 20.1) (Table~\ref{tab:demographics_combined}). Each participant selected four reproductive health-related articles, resulting in 136 article-based interactions. We recruited participants through social media, mass emails, and our social networks. This research received approval from our university's Institutional Review Board.

\begin{table}[t]
\centering
\caption{Demographics of the participants (N=34)}
\label{tab:demographics_combined}

\begin{tabular}{lcc}
\toprule
\textbf{Characteristic} & \textbf{Category} & \textbf{N (\%)} \\
\midrule
\multirow{2}{*}{Gender} & Female & 27 (79\%) \\
                        & Male & 7 (21\%) \\
\midrule
\multirow{4}{*}{Age (years)} & 18--19 & 20 (59\%) \\
                             & 20--22 & 10 (29\%) \\
                             & 25--26 & 3 (9\%) \\
                             & 29 & 1 (3\%) \\
                             & Mean (SD) & 20.1 (2.6) \\
\midrule
\multirow{6}{*}{Race/ethnicity} 
                      & Indian & 16 (47\%) \\
                      & Chinese & 8 (24\%) \\
                      & American & 7 (21\%) \\
                      & White & 1 (3\%) \\
                      & Norwegian & 1 (3\%) \\
                      & Sri Lankan & 1 (3\%) \\
\midrule
\multirow{6}{*}{Occupation} & Student & 26 (76\%) \\
                            & Software Engineer & 3 (9\%) \\
                            & Consultant & 2 (6\%) \\
                            & Product Manager & 1 (3\%) \\
                            & Sales Associate & 1 (3\%) \\
                            & Manager & 1 (3\%) \\
\bottomrule
\end{tabular}

\vspace{1mm}
\begin{minipage}{\linewidth}
\footnotesize
\textit{Note.} Race/ethnicity categories reflect participants' self-described responses.
\end{minipage}

\end{table}

\subsection{Data Analysis}
\begin{figure}[H]
    \centering
    \includegraphics[width=\linewidth]{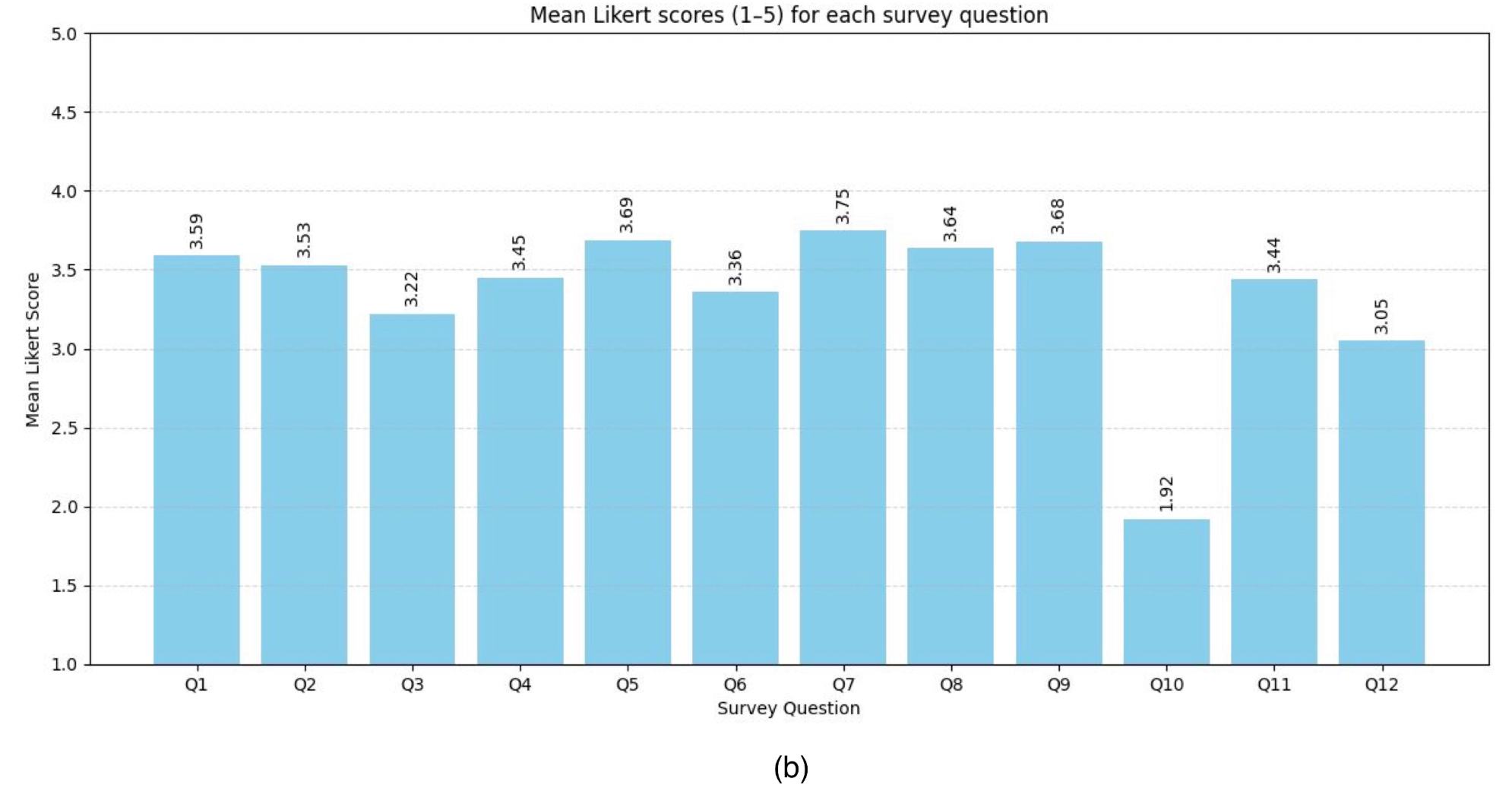}
    \caption{Mean Likert scores (1--5) for each survey question. Survey questions are listed in the Appendix. Q10 is negatively worded; lower scores indicate that participants generally did not find the questions offensive or inappropriate.}
    \Description{Bar chart showing the mean Likert score for each of twelve survey questions on a 1--5 scale. The mean scores are Q1: 3.59, Q2: 3.53, Q3: 3.22, Q4: 3.45, Q5: 3.69, Q6: 3.36, Q7: 3.75, Q8: 3.64, Q9: 3.68, Q10: 1.92, Q11: 3.44, and Q12: 3.05. Q7 has the highest mean score. Q10 is negatively worded and has the lowest mean, indicating that participants generally disagreed that the questions contained offensive or inappropriate language.}
    \label{fig:survey_graphs}
\end{figure}

We collected two forms of data following participants' interactions with OpenBloom: survey responses and semi-structured interviews. Participants first completed a quick survey containing twelve 5-point Likert-scale items (1 = Strongly Disagree, 5 = Strongly Agree) assessing participants' perceptions of the LLM-generated questions, including critical engagement, learning, relevance, comfort or safety, and cultural sensitivity. Given the exploratory nature of the study and the sample size ($N=34$), we conducted descriptive statistical analyses of the Likert-scale responses, reporting the mean score for each item (Figure~\ref{fig:survey_graphs}).

Following the survey, we conducted 20-30 minute semi-structured interviews to understand participants' experiences with the generated questions, with particular attention to cultural sensitivity, stigma, values, and opportunities for reflection. We analyzed the interview transcripts using thematic analysis \cite{braun2006thematic} to identify recurring patterns across participants' experiences and perceptions of OpenBloom.
\subsection{Results}

The overall results indicate participants' generally positive perceptions of the LLM-generated stigma-relevant questions across dimensions of critical engagement, learning, relevance, and cultural sensitivity. Across most items (Q1–Q9, Q11–Q12), responses were skewed toward Agree and Strongly Agree, with mean Likert scores clustering well above the neutral midpoint. This suggests that participants found the questions engaging, relevant, and thought-provoking in relation to reproductive well-being topics. In particular, high levels of agreement were observed for items concerning relevance, societal importance, and engagement. In contrast, Q10, which assessed whether any questions were perceived as offensive or inappropriate, showed a notably low mean score and a concentration of disagreement responses, indicating that most participants did not find the questions objectionable.

However, the follow-up semi-structured interviews painted a more nuanced picture of participants' perceptions, revealing limitations in the depth, relevance, and resonance with participants' lived experiences of the LLM-generated questions despite their generally safe and non-offensive nature. Several participants expressed concerns about the content of LLM-generated questions, including their lack of detail, limited critical depth, and tendency to resemble factual-recall questions. Five participants noted that the questions frequently rephrased sentences from the original article rather than encouraging deeper engagement. Additionally, four female participants noted small mismatches between the questions and the article content, including irrelevant details and inconsistent answer options in MCQ questions, which made the LLM-generated prompts feel more like an ``attention check''. They stressed that correctness alone was ``not enough'' for conversations involving sensitive reproductive issues. Also, many female participants connected the articles to reproductive stigma they had encountered in everyday life. One participant, for example, recalled her mother instructing her to wash her hair during menstruation because menstruation made a person ``dirty.'' In contrast, male participants generally found the questions more difficult to connect to their own lived experiences.

In terms of improvements, participants suggested that OpenBloom could be more contextual and culturally nuanced, such as by incorporating scenarios in questions. Moreover, questions can focus on meaningful content rather than trivial details. Besides MCQs, open-ended questions and multimedia elements (e.g., visuals or infographics) can be added to invite critical thinking and enhance engagement. While participants agreed on the importance of inclusive, empathetic language when discussing sensitive topics, some pointed out that what is considered “inclusive” cannot remain fixed as social understandings change: \textit{``What feels inclusive today might feel inadequate in five years.''} Notably, as mentioned above, participants emphasized that avoiding offensiveness alone was far from enough. Instead, they wanted systems to move beyond neutral, plausible, or ``safe'' language, and in some cases could include mildly provocative language to stimulate reflection and critical thinking. 
\section{Discussion}
Building on previous research on reproductive well-being, we designed OpenBloom, an LLM-mediated probe for reproductive well-being that encourages users to engage with reproductive health stigmas and reflect on their interpretations. Our findings from the initial survey and interviews raise both design and theoretical implications, which we discuss below.

\subsubsection*{Toward Contestability in Stigma-Sensitive Design}

Our findings contribute to feminist HCI and responsible AI scholarship by showing how stigma-sensitive AI interactions are shaped by gendered experiences, contextual sensitivity, and contestable values. Drawing on Feminist HCI~\cite{bardzell2010feminist}, the notion of ``the user'' can be updated to reflect gender in a way that ``noticeably and directly affects design.'' We observed that male participants generally found it more difficult to resonate with the themes of the AI-generated questions. These findings suggest that reproductive well-being technologies cannot assume a universal or neutral “user,” but instead must recognize how gender and social positioning shape engagement with sensitive health topics~\cite{kukura2022reconceiving}. Feminist HCI further highlights how evaluation can reveal “ways that designs configure users as gendered/social subjects”~\cite{bardzell2010feminist}. In OpenBloom, participants’ responses revealed how the system implicitly configured users through assumptions about relevance, experience resonance, and desirable ways to provoke reflections.

At the same time, our findings reveal that non-offensive and ``safe'' AI outputs are insufficient for stigma-sensitive design. While participants generally perceived the questions as non-offensive, several participants emphasized that overly neutral and superficial questions limited deeper engagement with reproductive stigma. Some even suggested that mildly provocative or challenging framings could better stimulate critical reflection. From a responsible AI perspective, these findings suggest the importance of designing for contestability, social repair, and harm mitigation, rather than focusing solely on the avoidance of offense~\cite{sultana2026deepfakes, yurrita2025contestability}.

This perspective also resonates with recent work in participatory AI justice, which argues that AI artefacts can function as forms of “adversarial design”~\cite{disalvo2012adversarial} that open up space for contestation and dissent and create agoras for constructive agonism~\cite{lupetti2026participatory}. Rather than positioning AI as a neutral authority, OpenBloom functioned as a design probe that surfaced disagreements, discomfort, and tensions around reproductive well-being discourse. Therefore, we argue that future stigma-sensitive LLM systems built for reproductive well-being and relevant domains~\cite{kim2025human, kelly2024chatbot} should support plural perspectives, situated contestability, and critical meaning-making, instead of merely optimizing for neutral or universally agreeable interactions.

\subsubsection*{Embracing Value-Sensitive Design in the Long Term}

The design literature in HCI has long examined value integration, including Value Sensitive Design (VSD)~\cite{friedman2019value}, values as lived experience~\cite{ledantec2009values}, value dilution in actions~\cite{Ghoshal2023dilution}, and values in repair~\cite{Repair}. Our participants shared that current LLM-generated questions defaulted to superficial rephrasing of source content and factual recall, limiting deeper reflection on stigmatized beliefs and practices. This finding extends prior work showing that stigma often persists even when accurate information is available~\cite{tuli2019girl, eshak2020myths}, and suggests that AI-mediated learning systems may similarly reproduce forms of “value dilution”~\cite{Ghoshal2023dilution}, where systems move away from the values they initially intend to support once enacted in practice. In our case, the values of reflection, empathy, and stigma-sensitive engagement became diluted into technically ``safe'' but socially shallow interactions.

Our findings further suggest the importance of moving beyond static or universal understandings of values in reproductive health technologies. Participants consistently preferred question framings that acknowledged lived experiences, emotional tensions, and sociocultural contexts, rather than relying solely on neutral informational correctness. This aligns with critiques of VSD that argue values should not only be predefined categories embedded into systems, but also discovered and negotiated through situated social contexts~\cite{ledantec2009values}. Participants particularly valued empathetic, scenario-based, and values-oriented prompts that encouraged critical reflection while recognizing existing beliefs. These findings point toward the need for shaping AI systems with moral imagination, where reproductive well-being technologies are designed to engage with the complexities, disagreements, and evolving meanings surrounding stigma-sensitive topics.

We highlight the importance of treating value-sensitive design as an ongoing sociotechnical process rather than a one-time design outcome. These limitations reinforce the need for long-term human-guided scaffolding, participatory refinement, and community accountability in stigma-sensitive AI systems. Consistent with Feminist HCI and participatory approaches, we argue that future reproductive well-being technologies should remain open to contestation, revision, and repair as values shift across contexts and over time.
\section{Limitations and Future Work}
Our participant sample was demographically skewed toward women, students, and participants from Indian and Chinese backgrounds. This limits the extent to which our findings can represent experiences across different age groups, occupations, genders, and cultural contexts. Future work will involve focus group discussions with a more demographically and culturally diverse set of participants to explore how users interpret, negotiate, and respond to LLM-generated questions across different cultural and social contexts. We also aim to iterate the system with provocative prompting strategies to explore how prompt questions affect users' reflections on reproductive well-being. These qualitative insights will inform the refinement of OpenBloom and contribute to more responsible, human-centered approaches to designing stigma-sensitive AI systems.



\bibliographystyle{ACM-Reference-Format}
\bibliography{citation.bib}

\appendix
\appendix

\section*{Appendix}

\section{Survey Questions}

Please indicate your level of agreement with the following statements about the questions generated from the article (1 = Strongly disagree, 5 = Strongly agree).

\begin{enumerate}
    \item These questions challenged me to think critically about reproductive well-being.
    \item These questions were interesting and engaging.
    \item These questions presented reproductive well-being topics in a unique and creative way.
    \item I resonated with the themes of the questions.
    \item These questions provided me with new knowledge or perspectives on reproductive well-being.
    \item These questions increased my awareness and motivation to challenge stigma around reproductive well-being.
    \item These questions accurately reflected the key themes of the article.
    \item These questions are relevant to social norms.
    \item These questions addressed issues that are important in today’s society.
    \item Some questions contained language or framing that I found potentially offensive or inappropriate.
    \item These questions made me feel comfortable discussing reproductive well-being topics.
    \item I found these questions to be sensitive to different cultural backgrounds.
\end{enumerate}

\section{Interview Guide}

\begin{enumerate}
    \setcounter{enumi}{0}
\item What were your overall impressions of the questions generated by OpenBloom? What made some questions feel more or less engaging or thought-provoking?

\item How did the questions relate to cultural or social norms around reproductive health?

\item In what ways did the questions align with or conflict with your values and lived experiences?

\item How did the wording or framing of the questions affect your willingness to respond openly? Were there moments when a question made you feel uncomfortable or defensive?

\item How did the questions engage with stigma, social norms, or taboos related to reproductive health? Can you give an example of a question that handled these issues well or poorly?

\item Did any questions encourage you to reflect on existing stigma or social norms around reproductive health?

\item How well did the questions describe factual information or provoke critical reflection?

\item Did you encounter any difficulties or issues while using the system?

\item What kinds of support would have helped you better understand or respond to the questions?

\item Do you have any suggestions for improving the system or the generated questions?

\end{enumerate}

\end{document}